\documentclass[%
 reprint,
 amsmath,amssymb,
 aps,
]{revtex4-2}

\usepackage{xcolor}
\usepackage{graphicx}
\usepackage{dcolumn}
\usepackage{bm}

\begin{document}

\preprint{APS/123-QED}

\title{Interactions of the solitons in periodic driven-dissipative systems supporting quasi-BIC states}

\author{D. Dolinina}
\email{d.dolinina@metalab.ifmo.ru}
\author{A. Yulin}%
\email{a.v.yulin@corp.ifmo.ru}
\affiliation{Faculty of Physics and Engineering, ITMO University, Saint Petersburg 197101, Russia}%


\date{\today}

\begin{abstract}
The paper is devoted to the dynamics of dissipative gap solitons in the periodically corrugated optical waveguides whose spectrum of linear excitations contains a mode that can be referred to as a quasi-Bound State in the Continuum.  
These systems can support a large variety of stable bright and dark dissipative solitons  that can interact with each other and with the inhomogeneities of the pump. One of the focus points of this work is the influence of slow variations of the pump on the behavior of the solitons. It is shown that for the fixed sets of parameters the effect of pump inhomogeneities on the solitons is not the same for the solitons of different kinds.
The second main goal of the paper is systematic studies of the interaction between the solitons of the same or of different kinds. It is demonstrated that various scenarios of inter-soliton interactions can occur: the solitons can repulse each other or get attracted.  In the latter case, the solitons can annihilate, fuse in a single soliton or form a new bound state  depending on the kinds of the interacting solitons and on the system parameters.
\end{abstract}

\maketitle


\section{INTRODUCTION}
In a vast realm of nonlinear optics, the formation and dynamics of the nonlinear periodic and localized patterns form a separate area attracting much attention for many years. The reason for this interest is the fundamental and practical importance of that dynamics in optical parametric oscillators, lasers, and other setups of such a kind. It is known that solitary waves known as gap or Bragg solitons can exist in these systems in the conservative and in the dissipative cases \cite{Mills, Mandelik, Sakaguchi, Yulin}. Let us remark that we use the term ``soliton'' in a wider sense meaning a nonlinear localized wave but not necessarily a solution of an integrable equation. 

The formation of the solitons in the conservative systems can be seen as the exact balance between the processes leading to the spreading (like, for instance, dispersion or diffraction) and the compression of the pulse (for example, a nonlinear focusing). In the case of dissipative systems, the energy balance must also be provided so that the losses are compensated by the pump. It should be mentioned here that dissipative solitons can have a conservative limit, but it is not a must. Normally the solitons are attractors in the sense that the shape of the solitons is determined by the parameters of the systems but is not sensitive to the small changes of the initial conditions \cite{Akhmediev}. The dissipative optical solitons have been found in fiber \cite{Renninger, Zhang_fiber, Peng_fiber} and semiconductor lasers \cite{Fedorov_semiconductor,Ultanir,Barbay_semiconductor}, optical parametric oscillators \cite{Oppo_OPO, Valcarcel_OPO, Babin},  high-Q ring resonators pumped by coherent light \cite{Herr_ring, Lucas_ring, Kippenberg}, and other systems.

One of the interesting systems where optical solitons can occur is that supporting so-called bound states in the continuum (BIC) \cite{BIC_review}. The BICs are the localized states having the frequency lying within the spectrum of the propagating waves. The BICs are of much interest in recent times because they open a possibility to achieve high-Q well-separated resonances. In the case of BIC, the Q-factor depends on internal losses only, because the radiative losses are exactly equal to zero. In real cases, the radiative losses are present but can be quite low, so it is possible to refer to these states as quasi-BIC states. High Q-factor allows BIC states to have high intensity of the field at a relatively low pump.  So, considering that optical nonlinearities are normally weak, the BIC and quasi-BIC states are very promising from the point of view of the experimental observation of nonlinear optical phenomena, see \cite{Koshelev_Sc,Koshelev} where highly efficient second harmonics generation is demonstrated experimentally.

The formation of the nonlinear switching waves (domain walls) and  dissipative solitons in driven-dissipative systems supporting  BIC is considered in \cite{Dolinina1, Dolinina2}. It is shown that in these systems there may exist bright solitons branching off the solitons in the conservative limit \cite{Dolinina1}. Another kind of dissipative solitons found in these systems is the bound states of the domain walls connecting different spatially uniform states \cite{Dolinina2}. Let us remark here that spontaneous symmetry breaking bifurcation can occur in the system and thus the solitons can nestle on the backgrounds with non-zero energy flow.

It is well known that dissipative solitons can be steered by the appropriate variation of the system parameters  \cite{Firth2, Rosanov_motion, Taranenko_motion, Maggipinto_motion}. From an experimental point of view, one of the most convenient control parameters is the pumping field because it allows dynamic control over the nonlinear patterns forming in the system. However, the dynamics of the solitons in the systems with other parameters slow varying in space can be of interest. 

If two solitons are separated by a relatively large distance then they can be treated perturbatively accounting for the inter-soliton interaction by the temporal variation of the soliton parameters. Indeed, in this case, the solution can be sought as a sum of two unperturbed solitons and a tail of the first soliton plays a role of weak perturbation seen by the second soliton. The interaction between the solitons can have different outcomes, for example, two solitons can annihilate \cite{Descalzi},  form soliton clusters \cite{Malomed, Schapers, Vladimirov},  demonstrate curvilinear motion \cite{Rosanov_int} or even chaotic behavior \cite{Turaev1,Turaev2}.

The purpose of the present paper is to study the motion of the solitons in driven-dissipative BIC systems under the action of the spatial variation of the system parameters and because of the inter-soliton interactions. 
The paper is structured as follows. In the next section, we discuss the physical system under consideration and introduce a mathematical model to describe the dynamics of the optical field. For sake of reading convenience we also briefly  summarize the main facts on the dissipative domain walls and the solitons existing in driven-dissipative BIC systems. In the third section we discuss the behaviour of dark and bright solitons in the pump field slow varying in time. We restrict our consideration to the cases when either the phase and or the amplitude of the pump field is varying in space. The fourth section of the paper is devoted to the various inter-soliton interactions taking place in the system. Finally, in the conclusion we briefly formulate the main findings reported in the paper.   

\section{THE PHYSICAL SYSTEM AND ITS MATHEMATICAL MODEL}

\begin{figure}[t]
\centering
\includegraphics[width=\linewidth]{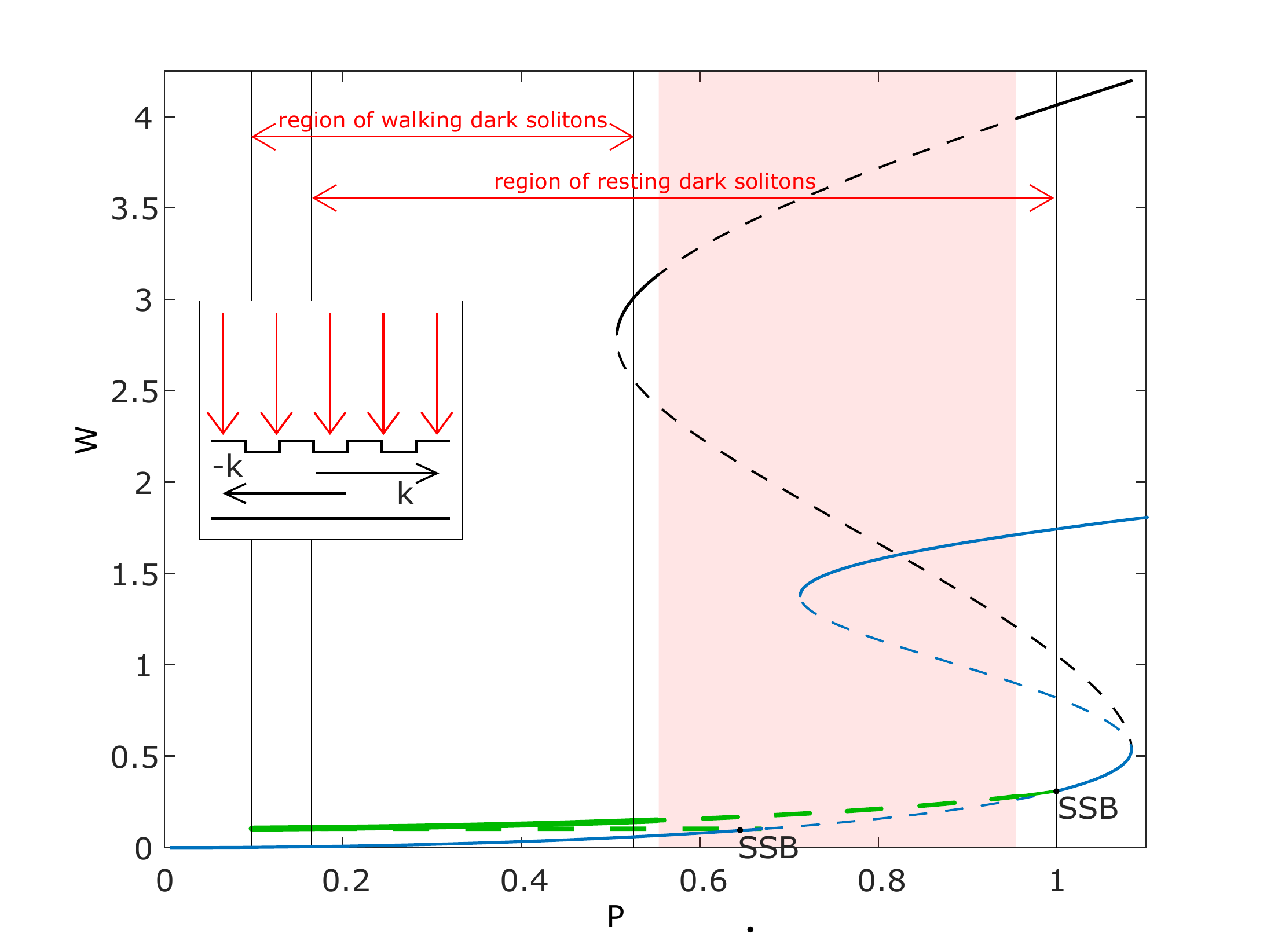}
\caption{Bifurcation diagram of the uniform states (blue line is for symmetric states and green line is for hybrid states) and bright solitons (black line), where $W = |U_b|^2 + |U_d|^2 = |U_+|^2 + |U_-|^2$ is intensity of states or maximum intensity of the solitons; dynamically unstable solutions are shown by dashed line. ``SSB'' points show the points where the spontaneous symmetry breaking bifurcation takes place; red rectangle stresses unstable uniform hybrid states and solitons; thin black vertical lines mark the region of existence of resting and walking dark solitons. Inset schematically shows the system under consideration. Parameters are: $\alpha = -1$, $\delta = 0.1155$, $\sigma = 1$, $\gamma = 0.001$, $\Gamma = 0.299$.}
\label{fig1}
\end{figure}

In the present paper, we consider a one-dimensional periodically grated waveguide with unfocusing nonlinearity of Kerr type under external laser pumping schematically shown in the inset of Fig.\ref{fig1}. This system was introduced earlier and studied in detail \cite{Krasikov2018,Dolinina1, Dolinina2}. 

Two coupled modes approach describing the light dynamics in the system can be written by the following equations \cite{Krasikov2018}:
\begin{eqnarray}
(\partial_t \pm \partial_x) U_{\pm}=&&(i\delta-\gamma) U_{\pm} + i\alpha (|U_{\pm}|^2 +2|U_{\mp}|^2)U_{\pm}+ \nonumber \\&&
+ (i\sigma - \Gamma)(U_{\pm}+U_{\mp}) +P,
\label{main_1}
\end{eqnarray}
where $U_{\pm}$ are the slowly varying amplitudes of waves propagating in positive and negative directions of the $x$ axis respectively, $\delta$ is the detuning of the pump frequency from the center of the gap of the dispersion characteristics, $\gamma$ accounts the internal losses, $\alpha = -1$ is the Kerr nonlinearity coefficient, $\sigma$ is the conservative part of the coupling coefficient defining the rate of the mutual re-scattering of the two counter-propagating waves, $\Gamma$ is the dissipative coupling accounting the fact that the radiative losses depend on the interference of the waves, and $P$ is the amplitude of the external coherent pump.

For further analysis it is convenient to re-write the equation in a different basis introducing new variables $U_b=\frac{U_{+}+U_{-}}{\sqrt{2}}$ and $U_d=\frac{U_{+}-U_{-}}{\sqrt{2}}$. The equations for $U_b$ and $U_d$ have the following form:
\begin{eqnarray}
(\partial_t -i\delta -i\dfrac{3}{2}K+ \gamma +2\Gamma - 2 i\sigma) U_b +(\partial_x  + &&iM)U_d= \nonumber\\=&&2P \label{bright}
\end{eqnarray}
\begin{eqnarray}
(\partial_t -i\delta -i\dfrac{3}{2}K+ \gamma) U_d +(\partial_x  + iM)U_b=0,
\label{dark}
\end{eqnarray}
where $K=\alpha(|U_b|^2+|U_d|^2)$, $M=\alpha Re (U_b U_d^{*})$. Let us note that in the linear case $\alpha = 0$ the spatially uniform field solution can be found in the form of the eigenmodes having the form $\vec U_b=(U_b \neq 0, U_d=0 )^T$ and $\vec U_d= (U_b=0, U_d \neq 0 )^T$.  The mode $\vec U_d$ experiences only internal losses (its radiative losses is absent ) but this mode cannot be excited directly by pumping, see Eq.(\ref{dark}). Therefore we refer it as ``dark'' or ``antisymmetric'' mode. The other mode $\vec U_b$ has nonzero radiative losses due to coupling to the freely propagating waves. This mode can be excited by the laser pumping at normal incidence so it will be referred as ``bright'' or ``symmetric'', see Eq.\ref{bright}.

Before further discussion, let us briefly reproduce the results from \cite{Dolinina1, Dolinina2}, in particular, the symmetry properties and dynamical stability of the uniform stationary states and dissipative solitons. Stationary spatially uniform states can be classified in three groups: symmetric states, which are purely bright modes; antisymmetric states, which are purely dark modes; and the combination of bright and dark modes that can be referred as the hybrid states \cite{Krasikov2018}. As it was shown in \cite{Dolinina1} the symmetric and hybrid states can be dynamically stable and thus can serve as backgrounds for the stable soliton and domain walls. The typical bifurcation diagram of the uniform states is shown in Fig.\ref{fig1}. The symmetric states are shown by the blue line and the hybrid ones by the green line and dynamical instability is shown by a dashed line. From Fig.\ref{fig1} it is clearly seen that part of the upper green branch is dynamically stable.

It is important to note that the hybrid states appear as a result of spontaneous symmetry breaking because of the parametric excitation of the dark mode. The hybrid states consist of two non-equal counter-propagating waves ($|U_+| \neq |U_-|$) and therefore its energy flux is nonzero $F_{HS}=|U_{+}|^2-|U_{-}|^2$. Because both directions are equivalent the bifurcation branches of hybrid states are double degenerated with states with the same absolute value of energy flux but of different signs.

In \cite{Dolinina2} it is demonstrated that such a system can provide domain walls connecting different spatially uniform states. The different domain walls can interact and form dark dissipative solitons as bound states. Such solitons can be classified as the resting solitons and the walking ones. The resting solitons are subdivided into two types: the solitons with energy flux directed towards the soliton and the solitons with energy flux directed away from the soliton. For further convenience, we will refer to the first type of resting dark solitons as ``sink-solitons'' and the later ones as ``source-solitons''. The region of existence of the dark solitons is shown in Fig.\ref{fig1} by red arrows.

Besides dark solitons, there are bright solitons in the studied system \cite{Dolinina1}. These solitons can be considered as a generalization of bright Nonlinear Schrödinger Equation solitons for the case of driven-dissipative systems. The typical bifurcation diagram of bright solitons is shown in Fig.\ref{fig1} by a black line.

The solitons of certain symmetries are resting if the pump is spatially uniform. However, in the non-uniform pumping field, the solitons are set in motion. We consider this motion in the next chapter. As is mentioned above there are also solitons that are moving even if neither the amplitude nor the phase of the pump depends on the coordinates. In the next chapter, we also study how the inhomogeneity of the pump affects the velocities of these solitons.

\section{MANIPULATING OF SOLITONS BY SPATIALLY NON-UNIFORM PUMPING}

\begin{figure}[t]
\centering
\includegraphics[width=\linewidth]{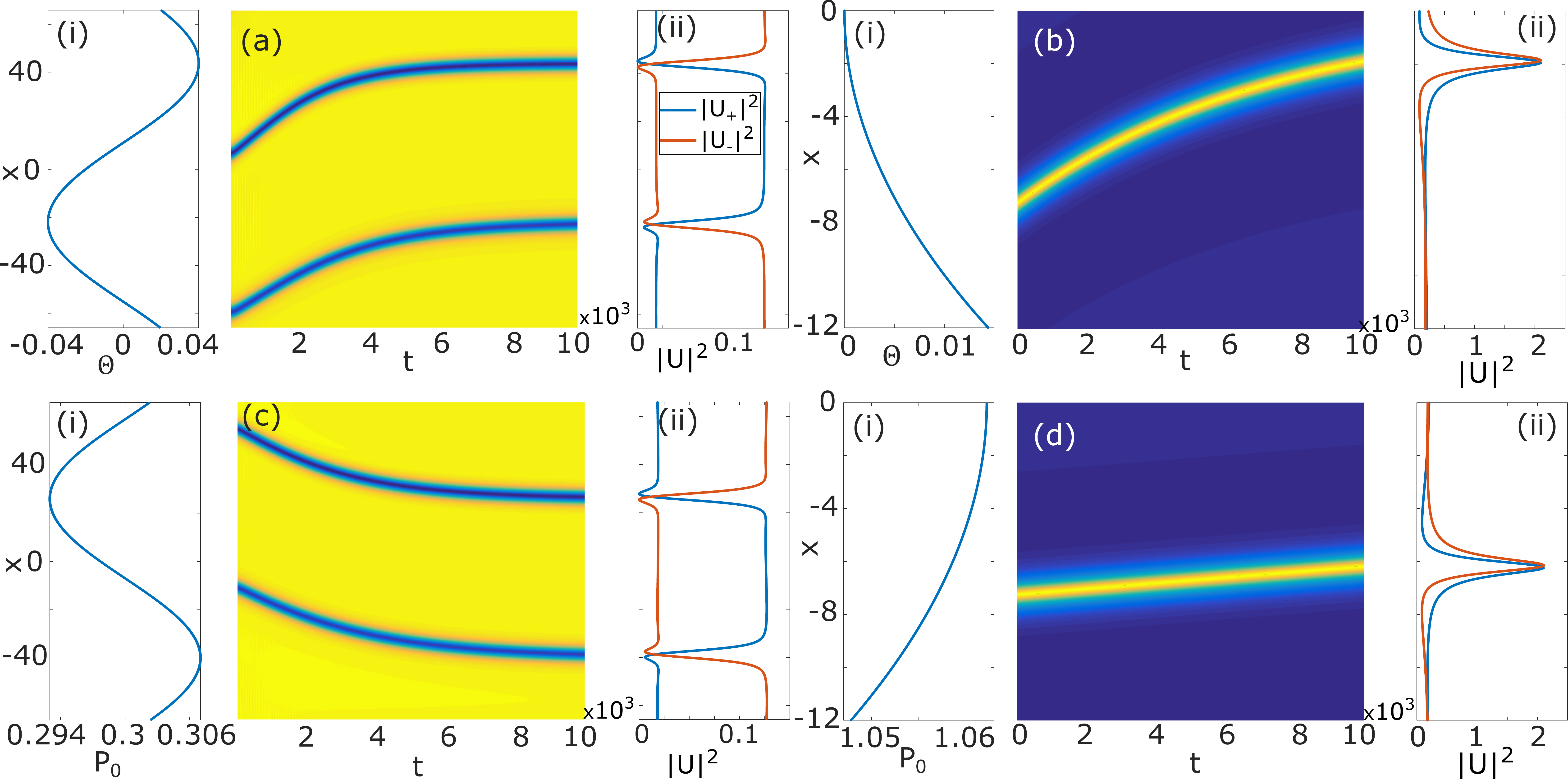}
\caption{Space-time diagrams demonstrating the motion of the solitons caused by spatially non-uniform  pumping. The colormap is used to show the intensity of the field $W(x,t)$ (yellow color corresponds to high field intensity and blue is for low intensity). Panels marked as ``(i)'' show modified phase (a-b) or amplitude (c-d) of the pumping field and panels ``(ii)'' show the final fields distributions. All parameters are the same as in Fig.\ref{fig1}. (a) Sink and source dark solitons moving under the pumping field with modified phase, $P = P_0 e^{\Theta(x)}$. (b) A bright soliton on the symmetric background set in motion by the pumping field with a non-uniform phase. (c) Amplitude gradient of the pumping field $P = P_0(x)$ forces sink and source dark solitons to move. (d) The bright soliton moving under the pumping field with non-uniform amplitude.}
\label{fig1a}
\end{figure}

As was mentioned before, the dissipative solitons can be set in motion by the phase or intensity gradient of the pumping field. In particular, it was shown that in the first approximation, the velocity of a dissipative soliton in a nonlinear optical cavity is proportional to the phase or intensity gradients of the pumping in the point of soliton location \cite{Firth2, Rosanov_motion, Taranenko_motion, Maggipinto_motion}. The relation between the effective force and the soliton velocity is expressed by so-called ``Aristotelian'' mechanics.

In the present paper, we study the soliton moving in the non-uniform pump by means of direct numerical simulations. Let us now briefly describe the numerical methods we used. First, we found stationary fields distributions of dissipative solitons under spatially uniform pumping field by solving stationary equations (\ref{main_1}) in a moving reference frame $\xi = x - Vt$ where the velocity $V$ was treated as an unknown variable. The obtained algebraic equations were solved iteratively by the Newton method. Then, these stationary fields distributions were used as initial conditions in direct numerical simulations with a slightly non-uniform pumping field. The simulations were made with the so-called split-step Fourier method, the essence of this method is that the linear part of the equation is solved in Fourier space where the spatial derivative operator is diagonal, and the nonlinear part is solved in coordinate space. This method is a very powerful and efficient tool in modeling nonlinear partial equations and it is widely used in optical studies.

Dynamically stable solitons are of the most interest from the physical point of view. Therefore, in the present paper, we limit ourselves to the discussion of the interaction between the stable solitons. It is important to note here that considered solitons nestling on stable backgrounds are always stable. The only exception is the dark solitons near Maxwell points where so-called ``snaking'' takes place and solitons of certain widths are unstable. Details of the snaking phenomenon in the system can be found in \cite{Dolinina2}. In other words, the existence of a stable background supporting dark dissipative solitons provides that some of these solitons are stable.

It is convenient to consider separately the case of the phase gradient and the gradient of the pump amplitude. Let us start with the case of solitons supported by the pumping field with the phase slowly varying in space $P = P_0 e^{\Theta(x)}$, see Fig.\ref{fig1a}(a,b). As is mentioned above the sink- and source dark solitons are not identical in terms of energy fluxes and so one can expect them to respond differently to the variation of the pump phase. Indeed, as it can be seen from Fig.\ref{fig1a}(a), the sink-soliton moves along the phase gradient while the source-soliton moves in the opposite direction. We also modeled the motion of the bright solitons under the action of the pump with spatially varying phase and find out that the bright solitons move in the direction opposite to the phase gradient, see Fig.\ref{fig1a}(b).

\begin{figure}[t]
\centering
\includegraphics[width=\linewidth]{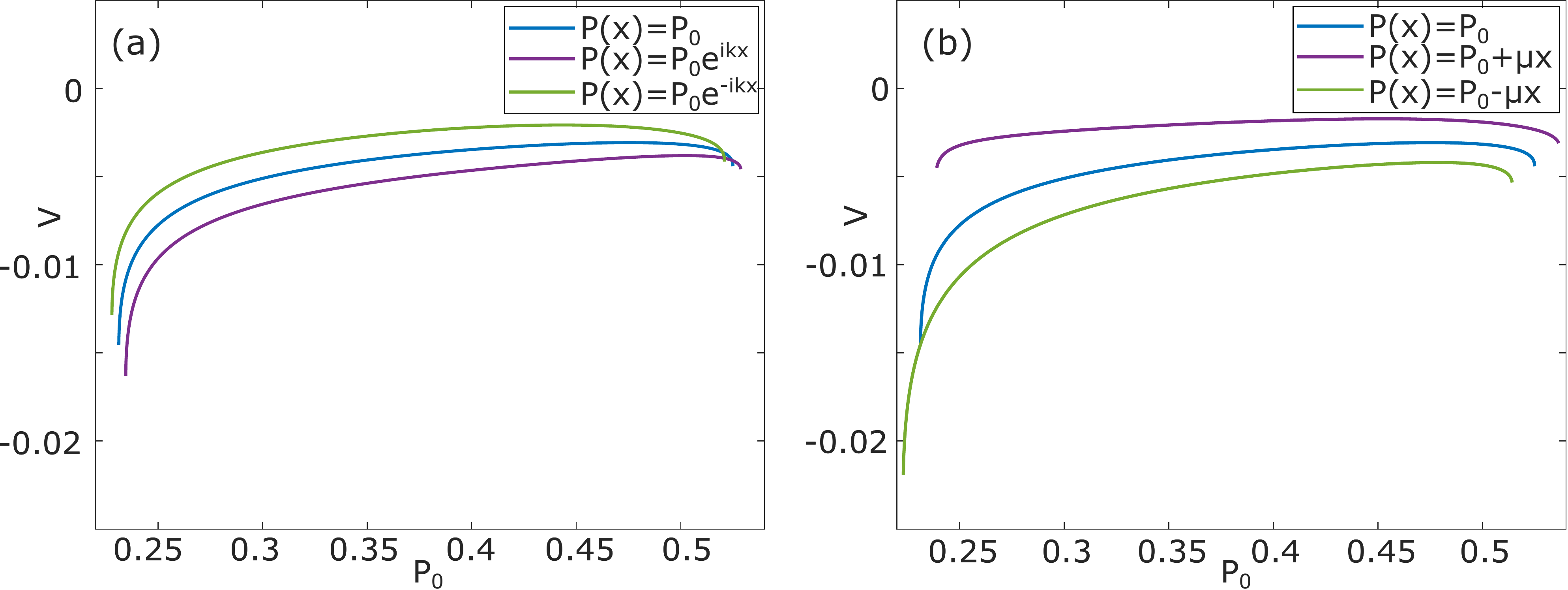}
\caption{Velocity dependence of the walking soliton at pumping amplitude. (a) Blue line shows velocity dependence in the uniform pumping field $P(x) = P_0$, violet one is for the pump field of the form $P(x) = P_0 e^{ikx}$ and green one is for $P(x) = P_0 e^{-ikx}$; $k = 10^{-3}$. (b) Blue line is the same as at (a), violet curve shows velocity dependence on pump of the form $P(x) = P_0 + \mu x$ and green one is for $P(x) = P_0 - \mu x$; $\mu = 5 \cdot 10^{-4}$.}
\label{velocity}
\end{figure}

The spatial variation of the pumping field amplitude $P = P_0(x)$ also sets the sink- and source solitons in motion, see Fig.\ref{fig1a}(c). From Fig.\ref{fig1a}(c) it is clearly seen that the sink-soliton moves toward the region with the lower pump intensity and the source-soliton oppositely gets pulled in the region of higher intensity. The bright solitons feel the variation of the pump intensity too, moving along the pump intensity gradient (to the region with the higher pump intensity), see Fig.\ref{fig1a}(d).

Now let us consider the behavior of the walking solitons in the system with a spatially non-uniform pumping field. The walking solitons exist in the background with spontaneously broken symmetry in such a way that the energy flows to the soliton from one side and flows away from the soliton on the other side. This explains why the solitons are  moving even if the pump is spatially uniform. Introducing a phase gradient of the pump it is possible to change the velocity of the walking solitons. We took the pump in the form $P=P_0 e^{ikx}$ and calculated the velocity of the walking soliton as a function of the pump amplitude $P_0$ for different phase gradients $k$. The results are presented in Fig.\ref{velocity}(a). One can see that the phase gradient of the pump affects the velocity of the solitons, the change of the soliton velocity is positive for the negative phase gradient and vice versa, the velocity shift is negative for the positive phase gradient. Let us remark here, that the effect of the pump phase gradient can be strong enough to change the sign of the velocity.

The effect of the intensity gradient on the soliton velocity is also studied for the pump taken in the form $P=P_0 + \mu x$ so that $\mu$ is the gradient of the amplitude of the pump. The dependencies of the soliton velocity on $P_0$ are shown in Fig.\ref{velocity}(b) for different gradients of the pump amplitude. In this case, the velocity also gets modified and can change its sign. It is found that the change of the velocity is of the same sign as the gradient of the pump amplitude. It should be noted here that we restricted our studies to the solitons belonging to the lowest stable branch of the soliton bifurcation diagram.

At the end of this section, we would like to remark that the bright dissipative solitons always behave in the same way as the dark source-soliton. Let us also point out here that the bright soliton on the hybrid background has energy flux directed towards the soliton. This means that the bright and the dark solitons with oppositely directed energy fluxes demonstrate qualitatively the same responses to the inhomogeneities of the pump.

\section{INTERACTION OF SOLITONS}

Let us now turn to the second problem considered in the paper and discuss how the dissipative solitons in the quasi-BIC systems interact with each other. All the presented results apply to the case of a spatially uniform pump. As the system can support solitons of different kinds, in particular, bright solitons on the background with unbroken and broken symmetry, resting source- and sink dark solitons, and walking dark solitons, we need to consider many cases of possible inter-soliton interactions. So in this paper, we discuss all possible interactions between the mentioned solitons. We start with the interaction of the dark resting solitons, then continue to the interaction of the dark resting with walking solitons, and then proceed to only walking solitons. After that, we take into consideration bright solitons and discuss their mutual interaction and their interaction with the dark resting and walking solitons.

\begin{figure}[b]
\centering
\includegraphics[width=\linewidth]{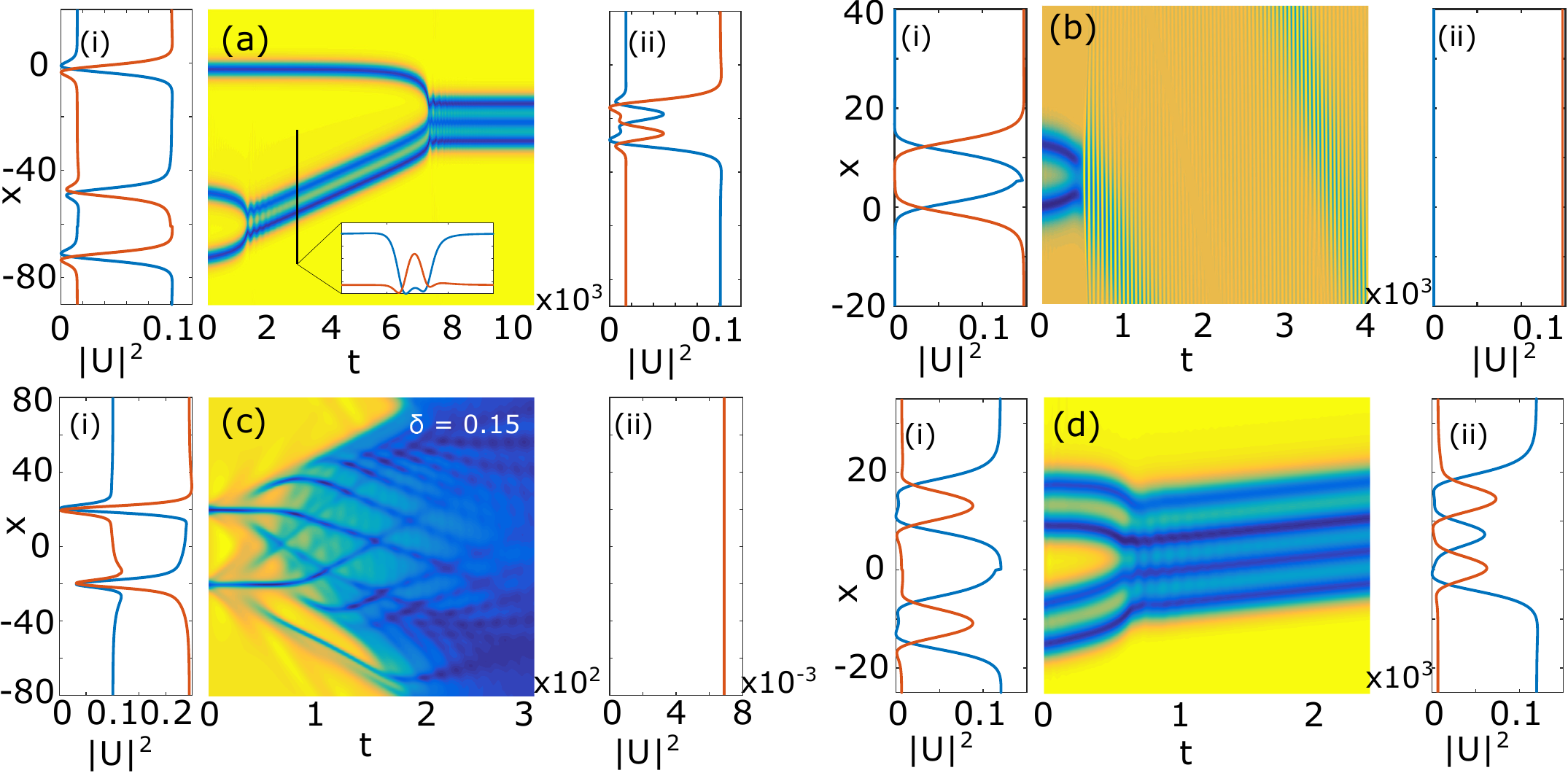}
\caption{Space-time diagrams demonstrating interactions of dark solitons. Panels (i) and (ii) show initial and final fields distributions, respectively. (a) Interaction of initially resting dark solitons resulting in walking soliton formation (see the inset) and further interaction of the walking soliton with the resting dark one, $P = 0.3$. (b) Interaction of dark solitons resulting in uniform hybrid state formation, $P = 0.54$. (c) Interaction of dark solitons resulting in uniform symmetric state formation; $\delta = 0.15$, $P = 0.3$. (d) Interaction of two walking solitons resulting in the formation of new walking soliton with a larger width, $P = 0.4$. Other parameters are the same as in Fig.\ref{fig1}.}
\label{fig2}
\end{figure}

Let us start with the interaction between dark solitons. As the initial condition, we take a combination of numerically found individual sink and source dark solitons. It is convenient to take three dark solitons, two of them are separated by a relatively short distant comparable to the size of the solitons. The third soliton is situated away from other solitons so that direct interaction between them is very weak. Let us remark that the solitons nestle on the backgrounds with nonzero energy flows and one can easily see that a neighbor of a sink-soliton must be a source-soliton, and, vice versa, the neighbor of a source-soliton must be a sink-soliton.

The initial distribution and the evolution of the resting dark solitons are shown in Fig.\ref{fig2}(a). One can see that a pair of the closely placed sink and source solitons get attracted to each other and after some time they collide. The collision results in the formation of a walking soliton, see \cite{Dolinina2} for details. The direction of the motion of the formed walking soliton depends on the relative positions of the sink and the source solitons in the initial conditions. Indeed, if the coordinate of a sink-soliton is less than the coordinate of the source soliton then after their collision the energy flow in the background of the resulting walking soliton will be directed along the $x$ axis. So the walking soliton will move along the axis. If the sink and the source solitons are swapped the walking soliton will propagate in the opposite direction.

In the case illustrated in  Fig.\ref{fig2}(a) the walking soliton moves towards the third resting soliton. Thus after some time they collide and this collision results in the formation of another resting soliton. This soliton can be seen as a bound state of two sink and one source solitons. Taking a different combination of the initial solitons we can obtain the formation of a resting soliton being a combination of two source and one resting soliton. So we can draw the following conclusions. First, the resting sink and source dark solitons get attracted to each other and collide. The collision can result in the formation of a walking soliton. Secondly, the walking soliton can collide with a resting dark soliton and form another resting soliton being, in fact, a bound state of three dark solitons of different kinds. 

However, the collision of a sink and a source solitons can have a different outcome when instead of the formation of a walking soliton the colliding solitons annihilate. The possibility of the annihilation follows directly from the fact that the ranges of the existence of the resting and walking solitons do not coincide and thus there is a range of parameters where the formation of the walking soliton is impossible. The evolution of dark solitons resulting in their annihilation and the formation of a spatially uniform state is shown in Fig.\ref{fig2}(b). The direction of the energy flow in the resulting uniform state is defined by the relative position of the sink- and source dark solitons.

If the detuning $\delta$ is relatively large then the interaction of the dark solitons causes perturbations so strong that they destroy the backgrounds of the solitons, see Fig.\ref{fig2}(c) switching the system to the lowest spatially uniform state. The collision causes strong perturbation of the solitons background and after a relatively long transitional process, the field becomes homogeneous in the whole system, see Fig.\ref{fig2}(c).

Our modeling shows that the interaction between the walking solitons is also attractive and thus they can collide, see Fig.\ref{fig2}(d) illustrating the dynamics of the field in this case. As a possible result of the interaction, the walking solitons form another walking soliton which is wider than each of the colliding solitons. Let us remark that if the initial solitons are of different widths and thus have different velocities, the collision introduces stronger perturbations so that the width of the final state is not approximately equal to the sum of the widths of the colliding solitons.

Let us note that for some sets of parameters the dark and the bright solitons can exist in the system simultaneously and thus they can interact. Firstly  we consider the interaction of two bright solitons. Two bright solitons can interact only if they nestle on a symmetric background with zero energy flux (in the case of a bright soliton on an asymmetric background the energy flow has to be pointed to the solitons in both the left and the right backgrounds and thus these bright solitons have to be separated by a domain wall).

\begin{figure}[t]
\centering
\includegraphics[width=\linewidth]{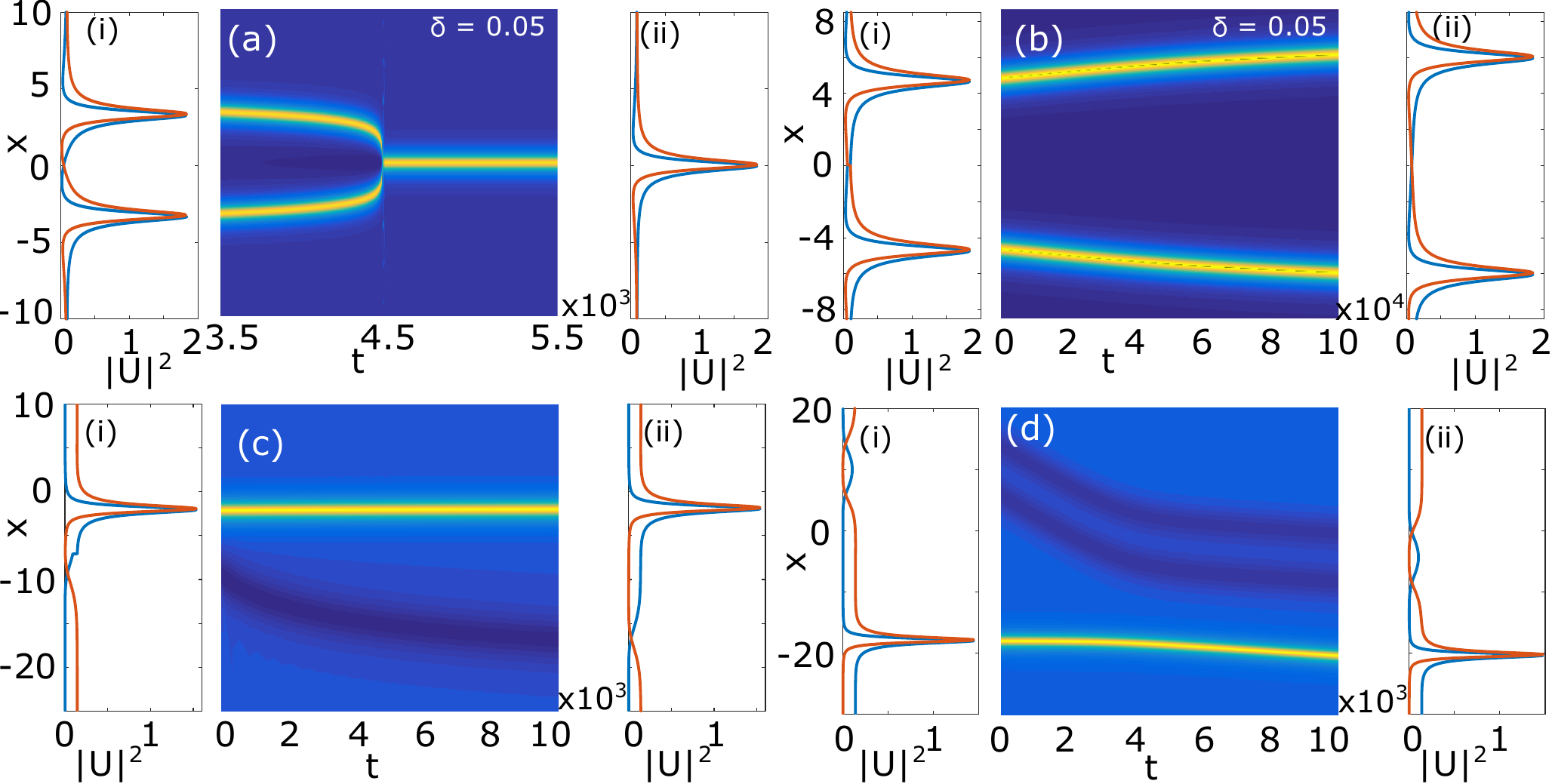}
\caption{Space-time diagrams demonstrating interactions of solitons. The colormap is used to show the intensity of the field $W(x,t)$ in (a)-(b) and the amplitude $\sqrt{W(x,t)}$ in (c)-(d). Detuning for (a)-(b) is $\delta = 0.05$, other parameters are the same as in Fig.\ref{fig1}. Panels (i) and (ii) show initial and final fields distributions, respectively. (a) Two bright solitons at the symmetric backgrounds collide and form a single soliton. Initial distance between solitons is $l = 7.9$. (b) Two bright solitons repulsing each other with initial dostance $l = 9.4$. For (a)-(b) $P = 0.8$. (c) Repulsive interaction of a source dark and a bright solitons, $P = 0.5244$. (d) Interaction of a walking soliton with a bright one resulting in formation of the moving bound state, $P = 0.51$.}
\label{fig3}
\end{figure}

It is interesting to note that in the considered system the interaction between the bright solitons can be either attracting or repulsing depending on the parameters of the system (the detuning $\delta$ and the pump intensity $P$) and on the distance between the interacting solitons. For relatively small detunings and pumping amplitudes (for instance, $\delta = 0.05$ and $P = 0.8$) the interaction is attractive if the distance between the solitons is relatively short. The case of attractive interaction between solitons initially placed at distance $l = 7.9$ is illustrated in Fig.\ref{fig3}(a). One can see that the collision results in a fusion of the solitons  into a single bright soliton.

However if the initial distance between two solitons is large enough then the interaction becomes repulsive and so the inter-soliton distance grows with time. Since the interaction strength depends on the inter-soliton distance exponentially, the interaction between the remote solitons is negligible. This kind of interaction is shown in  Fig.\ref{fig3}(b) where one can see how two solitons initially placed at distance $l = 9.4$ go away from each other but after some propagation time their velocities drop down dramatically.

An interesting problem is the interaction of a bright and a dark soliton. Let us discuss the interaction of a bright soliton and a resting dark solitons nestling on  asymmetric backgrounds. Let us remind that this interaction is possible only with dark source-soliton (because in the background separating the bright and the dark solitons the energy flow has to be directed towards the bright soliton).  The interaction is repulsive, see Fig.\ref{fig3}(b). Let us remark, the energy of the bright soliton is much larger compared to the energy of the dark soliton and this can be the reason why the motion of the bright soliton is hardly noticeable. So it looks like that during the interaction the  dark soliton runs away from its bright counterpart.

Finally, let us consider an interaction between a bright soliton and a walking dark soliton. The structure of the backgrounds is such that the walking soliton always moves towards the resting bright soliton. After the collision, a bound state of the bright and the dark state forms, see Fig.\ref{fig3}(d). The final complex state can also be considered as a walking soliton. These complex solitons can be of separate interest but their detailed investigation is out of the scope of the present paper.     

\section{CONCLUSION}

In this section, we briefly summarize and structure the effects found in the driven-dissipative nonlinear corrugated waveguides described within the framework of the two resonantly interacting counterpropagating waves. 

It is known that in such systems there may exist spatially uniform  states that can serve as backgrounds for different bright and dark dissipative solitons. In the paper, it is studied how these solitons behave if the pump varies slowly in space. In particular, we consider the case when the amplitude of the pump is a constant but the phase is a function of the coordinate. Another case considered in the paper is a pure real pump with the intensity varying in space.

It is shown that the inhomogeneities of the pump  affect not only the shape of the solitons but, more importantly, the velocity of the solitons. In particular, resting solitons of all kinds are set in motion by the pump inhomogeneities. The walking solitons are also studied and it is demonstrated that their velocities depend on the gradient of the pump. 

An important fact is that the interaction of the solitons with the pump inhomogeneity depends on the kind of the solitons. For example,  the sink dark solitons move along the phase and against the amplitude gradient. At the same time, the source dark solitons  move against the phase and along the amplitude gradient. 
This behavior opens a possibility to use the spatially non-uniform pump to make the sink and source solitons moving towards each other and collide. This is an example of how a spatially nonuniform pump can be used to manipulate the solitons. The systematic studies of the behavior of the solitons under the action of the non-uniform pump reported in the present paper can be used to organize the efficient steering of the solitons by the external coherent pump.  

Another problem considered in the paper is the interaction between the solitons. A large variety of the different solitons existing in the system makes it possible to observe multifarious scenarios of inter-soliton interaction.
In particular, it is shown that the interaction between resting dark solitons can result either in the formation of a moving bound state or in the annihilation of the solitons. In its turn, the collision of the walking dark soliton with a resting one produces a resting soliton. Thus way the walking solitons can be stopped by their resting counterparts.

Another important finding reported in the paper is that the interaction of two bright solitons on the symmetric background can be both attractive and repulsive depending on the system parameters, the intensity of the pumping field, and the distance between the solitons. 
It is also worth mentioning that in the considered systems there may be observed the interaction between the bright and the dark solitons. The collision of a walking dark soliton with a resting bright soliton gives birth to a complex moving soliton that can be seen as a stable bound state of bright and dark solitons.

We believe that the reported effects are not only of pure fundamental interest but possibly can have practical importance for the control of the dissipative solitons.  This is especially so considering that the systems with BIC allow to reduce significantly  the pump intensity needed to observe nonlinear effects. This could greatly facilitate the observation of the dissipative solitons in experiments. 

\begin{acknowledgments}
The work has been supported by Russian Fund for Basic Research (Grant ``Aspiranty'' No. 20-32-90227).
\end{acknowledgments}

\nocite{*}

\bibliography{apssamp}

\end{document}